\documentclass[12pt,english]{article}
\usepackage[latin1]{inputenc}
\usepackage{color}
\makeatletter

\font\mybb=msbm10 at 12pt
\def\bb#1{\hbox{\mybb#1}}

\usepackage{babel}
\makeatother
\begin{document}
%%%%%%%%%%%%%%%% title page %%%%%%%%%%%%%%%%%%%%%%%%%%%%%%%%%%%%

%%%%%%%%%%%%%%%% title page %%%%%%%%%%%%%%%%%%%%%%%%%%%%%%%%%%
\begin{titlepage}

\begin{flushright}
[hep-th]arXiv:0808.3568
%\\ 23/08/2008
%$\hspace{2.1cm}{}$
\end{flushright}

\vfill

\begin{center}
\baselineskip=16pt {\LARGE\bf NB BLG model in N=8 superfields}

\vskip 2cm

 {\large\bf Igor A. Bandos} \vskip 1cm {\small  Department of
Theoretical Physics and History of Science, The Basque Country
University (EHU/UPV), P.O. Box 644, 48080 Bilbao, Spain
 \\} \vspace{12pt}
{\small\it and \\} \vspace{12pt}
{\small  Institute for Theoretical Physcs, \\
NSC Kharkov Institute of Physics \& Technology, \\
UA 61108,  Kharkov, Ukraine. } \vspace{12pt}

\end{center}
\vfill

\par
\begin{center}
{\bf ABSTRACT}
\end{center}
\begin{quote}

We develop the N=8 superfield description of the
Bagger--Lambert--Gustavsson (BLG) model in its Nambu bracket (NB)
realization.

{\small

The basic ingredient is the octet of scalar $d=3$ $N=8$ superfields
$\phi^I$ depending also on the coordinates of a compact three
dimensional space $M_3$. It is restricted by the {\it
superembedding-like}  equation,
$\bb{D}_{\dot{A}}\phi^I=2i\psi_B\gamma^I_{B\dot{A}}$, which can be
treated as covariantization of the linearized superembedding
equation for supermembrane (M2-brane) with respect volume preserving
diffeomorphisms of $M_3$. The curvatures of SDiff$_3$ connection are
expressed through $\phi^I$ by the N=8 superfield generalization of
the BLG Chern--Simons equation (super-CS equation). We show how the
dynamical BLG equations appear when studying consistency of these
basic equations. }

\vfill \vfill \vfill \vfill \vfill \hrule width 5.cm \vskip 2.mm {\small \noindent
igor\_bandos@ehu.es; supported by the Basque Science Foundation {\it Ikerbasque}. }
\end{quote}
\end{titlepage}
%%%%%%%%%%%%%%%%%%%%%%%%%%%%%%%%%%%%%%

\section{Introduction}
\setcounter{equation}{0}

Recently, motivated by the search for the Lagrangian of multiple
M2-brane system, Bagger, Lambert \cite{Bagger:2007jr} and Gustavsson
\cite{Gustavsson:2007vu} proposed the $d=3$ $N=8$ supersymmetric
action based on Filippov 3-algebra instead of Lie algebra. A
particular infinite dimensional  3-algebra related with three
dimensional volume preserving diffeomorphism group SDiff$_3$ is
given in terms of Nambu brackets (Nambu--Poisson brackets)
\cite{Nambu} on a three dimensional compact manifold $M_3$. For
three functions on $M_3$, $\Phi(y^i)$, $\Xi(y^i)$, $\Omega(y^i)$
($i=1,2,3$), the Nambu brackets are defined by
\begin{equation}\label{NB}
\{ \Phi\, , \Xi \, , \Omega\}:= \bar{e}^{-1} \epsilon^{ijk}
\partial_i \Phi\,
\partial_j \Xi \, \partial_k \Omega\; , \qquad \partial_i:={\partial\over \partial
y^i}\;  \qquad
\end{equation}
(see \cite{Ho:2008nn,CornM5mM2} and refs. therein; here, following
\cite{IB+PKT=08,IB+PKT=08-2}, we have introduced a fixed $M_3$
density $\bar{e}=\bar{e}(y)$ in the definition of Nambu brackets).

As it was stressed in \cite{Ho:2008nn}, Bagger--Lambert--Gustavsson
model with Nambu bracket realization of the 3-algebra (NB BLG model)
can be treated as 6-dimensional field theory. The BLG gauge fields
become the gauge fields for the 3--volume preserving diffeomorphisms
(see \cite{Bergshoeff:1988hw} as well as \cite{IB+PKT=08-2} and
refs. therein). The SDiff$_3$ gauge potential is given by the 1-form
$s^i=dx^\mu s_\mu^i$ on $\bf{R}^{1+2}$ obeying the conditions
$\partial_i(\bar{e}s^i)=0$. The SDiff$_3$ field strength $F^i:=ds^i
+s^j\wedge \partial_js^i$ is also $M_3$ divergenceless,
$\partial_i(\bar{e}F^i)=0$,
\begin{equation}\label{diFi=0}
F^i:=ds^i +s^j\wedge \partial_js^i \; , \qquad
\partial_i(\bar{e}F^i)=0 \qquad \Leftarrow \qquad \partial_i(\bar{e}s^i)=0
\end{equation}
(see \cite{IB+PKT=08-2} for more details).

Furthermore, the authors of \cite{Ho:2008nn} proposed the
identification of the NB BLG model with M-theory 5-brane
($M5$--brane, see \cite{Howe:1996yn} for equations of motion and
\cite{Bandos:1997ui} for the covariant action) with the worldvolume
chosen to be $\bb{R}^{1+2}\otimes M_3$. However, an attempt to
obtain the NB BLG model from light-cone M5-brane \cite{IB+PKT=08}
has resulted only in reproducing the Carrollian limit of the NB BLG
model. This suggests to study the NB realization of BLG model
separately, and this was the subject of
\cite{CornM5mM2,IB+PKT=08,IB+PKT=08-2,Bonelli:2008kh} and also of
the original papers \cite{Ho:2008nn}.

In this letter we present N=8 superfield description of the NB BLG
model. It is the  on-shell superfield description which does not
allow for constructing the action, but reproduces equations of
motion as the selfconsistency conditions of the basic equations.

\section{Basic superfield equations}
\setcounter{equation}{0}

\subsection{Superembedding-like equation for octet of d=3 N=8 scalar superfields}

The complete on-shell N=8 superfield description of the Nambu
bracket realization of the Bagger-Lambert-Gustavsson model (NB BLG
model) is provided by the octet of scalar d$=$3, N$=$8 superfields,
$\phi^I=\phi^I(x^\mu\, , \theta^{\check{\alpha}}\, ;  y^i)$,
depending on additional coordinates $y^i$ ($i=1,2,3$) of a compact
space $M_3$, which obeys the following basic equation
\begin{equation}\label{DX=rp}
\bb{D}_{\alpha \dot{A}} \phi^I= 2i \tilde{\gamma}^I_{\dot{A}B}
\psi_{\alpha B}\; . \qquad
\end{equation}
Here and below $\alpha, \beta , \gamma=1,2$ are spinorial and $a,b,c
=0,1,2$ are vector indices of $SO(1,2)$,
$\tilde{\gamma}^I_{\dot{A}B}:= {\gamma}^I_{B\dot{A}}$ are the SO(8)
Klebsh-Gordan coefficients relating {\bf 8}$_v$, {\bf 8}$_s$ and
{\bf 8}$_c$ representation, which obey
$\gamma^{(I}\tilde{\gamma}{}^{J)}= \delta^{IJ}\, I_s$,
$\tilde{\gamma}{}^{(I}{\gamma}^{J)}= \delta^{IJ}\, I_c\;$
\footnote{The $SO(8)$ generators acting on {\bf 8}$_s$ and {\bf
8}$_c$ spinors are $ \gamma^{IJ}_{AB}\  :=
(\gamma^{[I}\tilde\gamma^{J]})_{AB}$ and $ \tilde\gamma^{IJ}_{\dot
A\dot B}  :=
 (\tilde\gamma^{[I}\gamma^{J]})_{\dot A\dot
 B}$.
Among the useful properties of these d=8 $\gamma$-matrices are
$\gamma^{{I}}_{A\dot A}\gamma^{{I}}_{B \dot B} = \delta_{AB}
\delta_{\dot{A}\dot{B}} + {1\over 4} \gamma^{{I}{J}}_{AB}
\tilde{\gamma}^{{I}{J}}_{\dot{A}\dot{B}}$,
$\;\gamma^{{I}{J}}\gamma^{{K}{L}}= \gamma^{{I}{J}{K}{L}} + 4
\delta^{[{I}|[{K}}\gamma^{{L}]|{J}]} - 2
\delta^{{I}[{K}}\delta^{{L}]{J}}$ and $\;\;
\gamma^{IJKL}_{AB}\gamma^{IJKL}_{\dot{C}\dot{D}}=0$. }, and
$\psi_{\alpha B}$ is a fermionic superfield which is expressed
through $\phi^I$ by the $\gamma^I$-trace part of Eq. (\ref{DX=rp}).
Finally $\bb{D}_{\alpha \dot{A}}$ is the covariant Grassmann
derivative. It is covariant with respect to $d=3$, $N=8$
supersymmetry and under the volume preserving diffeomorphisms of
M$_3$ (SDiff$_3$ group). Hence it involves a fermionic SDiff$_3$
connection $\varsigma^i_{\alpha \dot{A}} $ and, when act on
SDiff$_3$ scalars (like $\phi^I$ and $\psi_{\alpha B}$), reads
$\bb{D}_{\alpha \dot{A}} ={D}_{\alpha \dot{A}} + \varsigma^i_{\alpha
\dot{A}}\partial_i $ with ${D}_{\alpha \dot{A}}={\partial \over
\partial \theta^{\alpha}_{ \dot{A}}} +
i\gamma^\mu_{\alpha\beta}\theta^{\beta}_{ \dot{A}}\partial_\mu\;$ ($
\partial_\mu:={\partial \over
\partial x^\mu}$). These SDiff$_3$ covariant derivatives obey\footnote{\label{ftn1}For
simplicity, in (\ref{DX=rp}) we presented the anticommutator applied
to SDiff$_3$ scalar; the general expression is $\{ \bb{D}_{\dot{A}}
, \bb{D}_{ \dot{B}}\} = 2i \gamma^\mu \delta_{\dot{A}\dot{B}} {\cal
D}_\mu +2i \epsilon\, {\cal L}_{_{W_{\!\dot{A}\dot{B}}}}$, where
$W_{\dot{A}\dot{B}}:=W_{\dot{A}\dot{B}}{}^i\partial_i$ and ${\cal
L}$ is Lie derivative.}
\begin{equation}\label{(D,D)=D+W}
\{ \bb{D}_{\alpha \dot{A}} , \bb{D}_{\beta \dot{B}}\} = 2i
\gamma^\mu_{\alpha\beta}\delta_{\dot{A}\dot{B}} {\cal D}_\mu + 2i
\epsilon_{\alpha\beta} W_{\dot{A}\dot{B}}{}^i\partial_i\; , \qquad
\end{equation}
where ${\cal D}_\mu$ is the vector covariant derivative, which reads
as ${\cal D}_\mu= \partial_\mu + is_\mu^i\partial_i$ when acting on
SDiff$_3$ scalars. It involves the vector SDiff$_3$ gauge potential
$s_\mu^i$ defined by $s^i=d\theta_{\dot{A}}^\alpha
\varsigma_{\alpha\dot{A}}+
(dx^\mu-id\theta_{{\dot{A}}}\gamma^\mu\theta_{{\dot{A}}})s_\mu^i$.
The matrices $\gamma^\mu_{\alpha\beta}$ in (\ref{(D,D)=D+W}) are
real and symmetric; they obey $\gamma^{(\mu}\tilde{\gamma}^{\nu)}=
\eta^{\mu\nu}\delta_{\alpha}{}^{\beta}$, where $\tilde{\gamma}^{\mu
\,\alpha\beta}=\epsilon^{\alpha\gamma}\epsilon^{\beta\delta}\gamma^\mu_{\gamma\delta}$,
with
$\epsilon^{\alpha\beta}=-\epsilon^{\beta\alpha}=-\epsilon_{\alpha\beta}=i\tau^2=antidiag
(1,-1)$, and $\eta^{\mu\nu}=diag(1,-1,-1)$, is the flat metric in
the $d=3$ spacetime.

Finally,  $W_{\dot{A}\dot{B}}{}^i$ is the basic  superfield strength
of the SDiff$_3$ gauge supermultiplet. This carries the indices of
{\bf 28} representation of $SO(8)$, {\it i.e.}
$W_{\dot{A}\dot{B}}{}^i=-W_{\dot{B}\dot{A}}{}^i$, and is a vector
field with respect to SDiff$_3$ gauge group.

\subsection{N=8 superfield generalization of the Chern--Simons gauge field equation}

We impose on $W_{\dot{B}\dot{A}}{}^i$ the superfield generalization
of the Chern-Simons field equation. This reads \footnote{An
interesting, although technical, question is whether/how it can be
obtained from the consistency of the scalar superfield equation
(\ref{DX=rp}) and constraints (\ref{(D,D)=D+W}). However, as far as
there is no hope to get an off shell superfield model with 16
supersymmetries, at least in the 'standard' superspace, (so that the
real question is whether the constraints result in Chern-Simons or
in the D=3 SYM equations) in this letter we impose the
super-Chern-Simons equation as a constraint. }
\begin{equation}\label{WdAdBi=}
 W_{\dot{A}\dot{B}}{}^i=
 \bar{e}^{-1}\epsilon^{ijk}\partial_i\phi^I\partial_j\phi^J\tilde{\gamma}^{IJ}_{\dot{A}\dot{B}}\;
 \; , \qquad
\end{equation}
  Notice that $W_{\dot{A}\dot{B}}{}^i$ in
(\ref{WdAdBi=}) automatically satisfies the condition
$\partial_i(\bar{e}W_{\dot{A}\dot{B}}{}^i)=0$ necessary for any
SDiff$_3$ field strength \cite{IB+PKT=08-2} (see (\ref{diFi=0})).

As far as $\tilde{\gamma}^{IJ}_{\dot{A}\dot{B}}$ form the complete
basis in the space of antisymmetric 8$\times$8 matrices, an
equivalent form of the super Chern--Simons equation (\ref{WdAdBi=})
is given by
\begin{equation}\label{WIJi=}
 W^{IJ \;i}=
 \bar{e}^{-1}\epsilon^{ijk}\partial_i\phi^I\partial_j\phi^J\;
 \; , \qquad  W_{\dot{A}\dot{B}}{}^i=: W^{IJ
 \;i}\tilde{\gamma}^{IJ}_{\dot{A}\dot{B}}\; . \qquad
\end{equation}

\section{Bagger--Lambert equations of motion from the basic
superfield equations}

The spinor covariant derivative 'algebra' (\ref{(D,D)=D+W})
simulates the constraints for SYM fields. However, Eq.
(\ref{WdAdBi=}) implies that the corresponding SDiff$_3$ gauge
theory supermultiplet is composed in the sense that all the field
strengths are expressed through the scalar and spinor fields.

Indeed, studying the consequences of the gauge field Bianchi
identities\footnote{i.e. Jacobi identities for the covariant
derivatives, $[\bb{D}_{\alpha \dot{A}}\, , \, \{ \bb{D}_{\beta
\dot{B}} \, , \, \bb{D}_{\gamma \dot{C}}\}] + \left(  ^{_{\alpha
\dot{A}}}\matrix{_\nearrow \,  ^{\beta \dot{B}}_{\; \downarrow} \cr
^{^\nwarrow} \, ^{\gamma \dot{C}}} \right)=0$ {\it etc.}} one finds,
firstly, that the commutator of vector and spinor covariant
derivatives reads ${}[\bb{D}_{\alpha\dot{B}},{\cal D}_a]=
i\gamma_{a\alpha\beta}W^{\beta \: i}_{\dot{B}}\partial_i $ and that
the Grassmann spinor octet field strength $W_{\alpha\dot{B}}{}^i $
is given by
\begin{equation}\label{WaldAi=}
W_{\alpha \dot{B}}{}^i= {i\over 7} {\bb D}_{\alpha \dot{A}}
W_{\dot{A}\dot{B}}{}^i = 4
 \bar{e}^{-1}\epsilon^{ijk}\partial_j\phi^J\partial_k\psi_{\alpha{A}}\,
{\gamma}{}^{J}_{{A}\dot{B}}\;
 \; , \qquad
\end{equation}
and, secondly, that tensorial gauge field strength (${}[{\cal
D}_\mu\, ,{\cal D}_\nu]= F_{\mu\nu}^i\partial_i $) reads
\footnote{These equations can be also obtained from consistency of
(\ref{DX=rp}) with the use of  (\ref{WdAdBi=}) (see below).}
$F_{\mu\nu}{}^i = -{1\over 16}\epsilon_{\mu\nu\rho}
\gamma^\rho_{\alpha\beta}\bb{D}^\alpha_{\dot{A}}W^{\beta\;
i}_{\dot{A}}$ so that
\begin{eqnarray}\label{CS=Eq}
&&  F_{\mu\nu}{}^i = - 2 \bar{e}^{-1}
\epsilon^{ijk}\epsilon_{\mu\nu\rho}\left(
\partial_j\phi^J\partial_k{\cal D}^{\rho}\phi^J +2i
\partial_j\psi_{A} \gamma^\rho\partial_k\psi_{A} \right)\; .
 \qquad
\end{eqnarray}
In (\ref{CS=Eq}) one recognizes the Chern--Simons type gauge field
equations which can be obtained from the BLG Lagrangian of
\cite{Bagger:2007jr}. This expresses the tensorial gauge field
strength through the matter (super)fields.

The dynamical bosonic and fermionic equations of motion of the NB
BLG model follow from the superembedding--like equation
(\ref{DX=rp}) and the super-CS equation (\ref{WdAdBi=}). Indeed,
with the use of (\ref{(D,D)=D+W}), one finds that the
selfconsistency condition for Eq. (\ref{DX=rp}) gives the expression
for Grassmann covariant derivative of the fermionic superfield
$\psi_{\beta B}$ in (\ref{DX=rp}),
\begin{eqnarray}\label{Dpsi=}
\bb{D}_{\alpha \dot{A}} \psi_{\beta B}&=& {1\over 2}
\gamma^\mu_{\alpha\beta} {\cal D}_\mu \phi^I {\gamma}^I_{B\dot{A}} +
{1\over 3! \,\bar{e}}\, \epsilon_{\alpha\beta}
 W^{IJ \;i}\partial_i\phi^K \tilde{\gamma}^{IJK}_{\dot{A}B} = \\
 &=& {1\over 2}
 {\cal D}\!\!\!\!/{}_{\alpha\beta} \phi^I
{\tilde\gamma}^I_{\dot{A}B} + {1\over 6} \, \epsilon_{\alpha\beta}\,
\{ \phi^I\, , \phi^J \, , \phi^K\} \,
\tilde{\gamma}^{IJK}_{\dot{A}B} \nonumber \; .\qquad
\end{eqnarray}
Next stage is to study the selfconsistency conditions for Eq.
(\ref{Dpsi=}). Its SO(1,2) vector and SO(8) tensor ($\propto
\tilde{\gamma}^{IJKL}\gamma^\mu_{\alpha\beta}$) irreducible part
gives us the expression (\ref{WaldAi=}) for the spinor field
strength of the SDiff$_3$ gauge field (fermionic superpartner of the
BLG Chern-Simons equation (\ref{CS=Eq}))\footnote{On the way of such
a derivation of (\ref{WaldAi=}) one should use the requirement of
that the dependence of $M_3$ coordinates should not be restricted,
i.e. no additional conditions on $\partial_i\phi^I$ may occur. Then,
coming to the equation $(W_{\dot{A}}^{\; i}-
\ldots)\partial_i\phi^I= \kappa_{B}\gamma^{I}_{B\dot{A}}$, one
concludes that $\kappa_{B}=0$ and that $W_{\dot{A}}^{\; i}= \ldots$
where multidots denote the {\it r.h.s} of Eq. (\ref{WaldAi=}).}.
Taking this into account in the SO(1,2) vector - SO(8) scalar
($\propto \delta_{\dot{A}\dot{B}}\gamma^\mu_{\alpha\beta}$)
irreducible part we obtain the BLG Dirac equation
\begin{eqnarray}\label{Dirac=Eq}
 \gamma^{\mu}_{\alpha\beta}{\cal D}_\mu \psi^\beta_B &=&
- {1\over  \bar{e}}\; \epsilon^{ijk} \partial_i \phi^I
\partial_j\phi^J \partial_k \psi_{\alpha A} \gamma^{IJ}_{AB}\;  ,
 \qquad
\end{eqnarray}
which can be equivalently written in the following compact form
\begin{eqnarray}\label{Dirac=Eq2}
 {\cal D}\!\!\!\!/\;{} \psi
=-\, \{ \phi^I \, , \phi^J \, , \psi \} \gamma^{IJ} \; .
 \qquad
\end{eqnarray}

 As usually, the bosonic equations of motion can
be obtained by taking the covariant spinorial derivative of the
fermionic ones. Acting by the covariant (SDiff$_3$ and SUSY
covariant) spinor derivatives on  (\ref{Dirac=Eq}), and extracting
the $\propto \epsilon_{\alpha\beta}\gamma^I_{C\dot{A}}$ irreducible
part one finds
\begin{eqnarray}\label{K-G=Eq}
 {\cal D}^\mu{\cal D}_\mu\phi^I = \, 2\,  \{ \phi^J \, , \phi^K \, ,  \, \{  \phi^I
 \, ,  \phi^J  \, , \phi^K  \}\} \, - \, 4i\; \epsilon^{\alpha\beta} \{  \psi_{\alpha}\, ,  \gamma^{IJ}
 \psi_{\beta} \, ,  \phi^J
 \, \}
  \;
 \qquad
\end{eqnarray}
The $\propto \gamma^\mu_{\alpha\beta}\gamma^I_{C\dot{A}}$
irreducible part of the same relation can be used to obtain the
bosonic Chern-Simons equation (\ref{CS=Eq}), while the $\propto
\gamma^{IJK}_{C\dot{A}}$ irreducible parts vanish
identically\footnote{To prove this,  one has to use the consequences
$\{\phi^L\, , \phi^{[I}\, , \,   \{\phi^J\, , \phi^{K]}\, , \phi^L\,
\}\} =0 $  and $\epsilon^{IJKLMNPQ}\{\phi^L\, , \phi^M\, , \,
 \{\phi^N\, , \phi^P\, , \phi^Q\, \}\} =0 $ of the so-called
fundamental identity $\{ \phi^L\, , \phi^M\, ,
 \, \bar{e} \{\phi^N\, , \phi^P\,
, \phi^Q\, \}\}= 3\{ \, \bar{e}\{\phi^L\, , \phi^M\, \phi^{[N}\, \}
, \phi^P\, , \phi^{Q]} \}\}$ (the presence of the density
$\bar{e}(y)$ in the fundamental identity and its absence in its
consequences above is not occasional).}.

To conclude, the superembedding--like equation (\ref{DX=rp}),
supplemented by the covariant derivative algebra (\ref{(D,D)=D+W})
with the composite scalar field strength (\ref{WdAdBi=}), restricts
field content of the basic octet of d=3, N=8 scalar superfields
$\phi^I$, depending in addition on three coordinates of a compact
space $M_3$, to the NB BLG supermultiplet, and, furthermore,
accumulates all the equations of motion of the NB BLG model.

\bigskip

\section{Conclusions}

In this letter we presented the N=8 superfield description of the
Nambu bracket (NB) realization of the Bagger--Lambert-Gustavsson
(BLG) model. It is given by an octet of scalar N=8, d=3 superfields
$\phi^I$ which, in addition, depend on the three coordinates $y^i$
of compact space $M_3$. This octet of superfields is restricted by
Eq. (\ref{DX=rp}), which, as we have shown, contains all the
equations of motion of the NB BLG model when supplemented (at least,
when supplemented) by super-Chern-Simons equation (\ref{WdAdBi=})
(or, equivalently, (\ref{WIJi=})).

We call the basic Eq. (\ref{DX=rp}) {\it superembedding--like
equation} because of its  relation with the superembedding equation
describing one M2-brane in the d=3 N=8 worldvolume superspace  which
is as follows. To obtain (\ref{DX=rp}), one has first to linearize
the supermembrane superembedding equation \cite{bpstv}, see
\cite{Howe:2004ib}, and to fix the so-called static gauge on the
worldvolume superspace, arriving at the equation $D_{\alpha
\dot{A}}X^I = 2i \tilde{\gamma}^I_{\dot{A}B} \Psi_{\alpha B}$.  Then
one replaces the octet of d=3, N=8 superfields $X^I(x,\theta)$ by
the octet of superfields depending also on coordinates of $M_3$,
$X^I(x,\theta)\mapsto \phi^I(x,\theta, y)$ (this automatically
produces $\Psi_{\alpha B}(x,\theta)\mapsto \psi_{\alpha B}(x,\theta,
y)$) {\it and} covariantize the result with respect to the volume
preserving diffeomorphisms of $M_3$ (${D}_{\alpha \dot{A}}\mapsto
\bb{D}_{\alpha \dot{A}} ={D}_{\alpha \dot{A}} + \varsigma^i_{\alpha
\dot{A}}\partial_i $).

We hope that our superfield description will be useful in studying
the properties of the NB BLG model  and in understanding its
physical meaning.

Actually, such a way of passing from the complete nonlinear
description of one M2-brane in the frame of superembedding approach
\cite{bpstv} to the NB BLG model-- namely first linearization, and
than obtaining nonlinearities by a separate covariantization with
respect to SDiff$_3$,-- suggests that NB BLG model may be not a
description of multiple M2-brane, but rather an independent- and
without any doubt very interesting- $d=3$, $N=8$ supersymmetric
dynamical system.

Actually, a search for alternative candidates on the r\^ole of
multiple M2-brane action can be witnessed. A very incomplete list
includes the $N=6$ supersymmetric model of \cite{Aharony:2008ug},
d=3, N=2 supersymmetric models of \cite{Cherkis:2008qr}, as well as
very recent construction of a candidate multiple M2-brane bosonic
action, similar to the Myers action for the coincident bosonic
D-branes, in \cite{Iengo:2008cq}.

The generalization of our on-shell N=8 superstring description for
the case of arbitrary 3--algebra seems to be possible\footnote{This
is better seen when one writes the SDiff$_3$ covariant derivatives
in terms of Lie brackets  of vector fields, $\bb{D}\phi^I=D\phi^I +
[\varsigma \, , \phi^I]$, $F=ds+ {1\over 2} [s , s ]$. Then these
Lie brackets can be substituted by the commutators and the
commutators of the field strengths are defined by 3-brackets with
scalar and spinor fields, {\it e.g.} ${}[W^{IJ}\, , \, ...\, ]=  \{
\phi^I\, , \phi^J \, , \, ...\, \} $.} and, in many respects, looks
interesting to develop in details.

\bigskip
\noindent
\section*{Acknowledgments}

The author is grateful to Paul Townsend for useful discussions and
for collaboration in studying the SDiff$_3$ gauge theories, and
thanks Neil Lambert for a discussion in CERN, on String 08
conference. This work was supported by the Basque Science Foundation
{\it Ikerbasque} and partially by research grants from the Spanish
MICI (FIS2008-1980), the INTAS (2006-7928), the Ukrainian National
Academy of Sciences and Russian RFFI grant 38/50--2008.

\bigskip

\section*{Notes added} When this paper has been finished, the
article \cite{Cederwall:2008vd}, addressing the same subject from
the pure spinor (pure spinor superspace) perspective, has appeared
on the net. A very intriguing statement in \cite{Cederwall:2008vd}
is on existence of the superspace action that, if so, would be the
first known example of the superfield action with 16
supersymmetries. The action presented in \cite{Cederwall:2008vd} can
also be considered as a realization of the harmonic superspace
programme of \cite{GIKOS} with pure spinors substituting harmonic
variables\footnote{As the pure spinors for $D=10$, $N=1$ superstring
\cite{Berkovits2000} parametrize, modulo overall scale factor, the
$SO(10)/U(5)$ coset, one can also state that these pure spinors {\it
are} harmonic variables.}. The approach of \cite{GIKOS} overcame
some no-go theorems because the number of auxiliary fields in it was
infinite. It would be very interesting to analyze the structure of
auxiliary field sector in the action of \cite{Cederwall:2008vd}.

After the first version of this paper appeared on the net, certain
aspects of measure on the pure spinor space, which were left out and
simply assumed to work in \cite{Cederwall:2008vd}, were addressed in
\cite{Cederwall:2008xu}, where a similar pure spinor superspace
formulation was also presented for the N=6 model of
\cite{Aharony:2008ug}.

 \end{document}